\documentclass[a4paper]{llncs}

\usepackage{hyperref}
\usepackage{eso-pic}
\AddToShipoutPicture*{
\put(110,80){
\scriptsize
Proceedings of \href{http://arxiv.org/abs/1406.1510}{11\textsuperscript{th} Workshop on Constraint Handling Rules (CHR 2014)}, 
paper CHR/2014/1.
}}

\usepackage{a4}
\usepackage{amssymb}
\usepackage{graphicx}
\usepackage{float}

\floatstyle{plain}
\newfloat{listing}{thp}{lst}
\floatname{listing}{Listing}

\begin{document}

\title{Automatic Test Data Generation and Model Checking with CHR}
\author{Ralf Gerlich}
\institute{BSSE, Auf dem Ruhbuehl 181, D-88090 Immenstaad, Germany,\\
\email{ralf.gerlich@bsse.biz}}
\maketitle

\begin{abstract}
We present an example for application of Constraint Handling Rules to automated test data generation and model checking in verification of mission critical software for satellite control.
\end{abstract}

\section{Overview}
Verification and validation of software takes up a large proportion of project effort and cost, especially in the area of mission and safety critical software.
This is one of the driving forces for automation of these aspects of software engineering, besides the reduction of the potential for errors due to manual labor.

Automation of software testing requires among others the automatic generation of test data.
Some options for this are random test data generation~\cite{chen.huang.ea2007,gerlich.fercher1993,gerlich2005c,hamlet1994} or constraint-based test-data generation (CBDTG)~\cite{denise.gaudel.ea2004,ferguson.korel1996,gerlich2010,godefroid.klarlund.ea2005,gotlieb.botella.ea1998,korel1990,nguyen.deville2001}.

We have designed and implemented a toolchain for CBTDG using Constraint Handling Rules (CHR), which we have already presented in the past~\cite{gerlich2010}.(in [13] and [4] the two examples of the paper are noted
  using typica

In a recent case-study on the effectiveness of source-code-based random testing~\cite{gerlich.gerlich.ea2013}, we have also seen that the basic elements of this toolchain can be used for support in manual verification of defects in mission critical satellite control software and that a use for model checking seems plausible.

\section{Introduction}
Constraint-based test data generation is concerned with the generation of program inputs for use in software test using constraint programming techniques.
The goal is to find program inputs which fulfill specific criteria, typically from structural coverage goals such as executing a specific portion of the program under test.

In a first step, in our approach -- similar to that of others -- the control flow graph of the function under test is used to construct a path that fulfills the request.
For each path an associated \emph{path constraint} can be constructed, which is actually a set of constraints.

\begin{figure}
\centering
\includegraphics{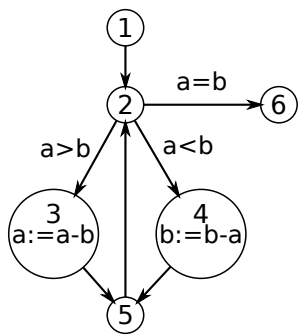}
\caption{Control Flow Graph}
\label{fig:cfg}
\end{figure}

An example control-flow graph for an implementation of Euclid's algorithm for determination of the greatest common divisor of two positive integers is given in Fig.~\ref{fig:cfg}.
Execution starts at node $1$ and continues along the edges until node $6$ is reached.
Nodes and edges are annotated by statements and conditions, respectively.
An edge may be traversed only if the condition attached to it is fulfilled.

The path constraint is a constraint over the input variables of the function, and its solutions are the inputs that fulfill the given criteria.
Thus, in a second step, a solution to the path constraint is sought.

There are paths in the control flow graph which are associated with a path constraint without solution.
These paths are called \emph{infeasible} and -- unfortunately -- may represent a large portion of all paths, so that a simple trial-and-error-approach to path construction is insufficient.

Instead, infeasible paths should be detected early in the path construction phase.
This requires both a useful strategy of path construction and a constraint solver which is efficient at detecting inconsistencies.

The criteria applied to the desired input can also be described in the form of constraints, allowing integration of the structural goals for the test input with, e.g., arithmetic constraints on the state of program variables at specific stations during execution.
This combination also makes the use of the same approach for static verification using symbolic enumeration of the state space possible, as will become apparent from our example.

This paper is structured as follows: In Section~\ref{sect:constraint_solver_approach} we detail the constraint solver used and the diverse requirements it must fulfill.
This is followed by a short discussion on the reasons for applying CHR for implementation in Section~\ref{sect:application_of_chr} and the presentation of open problems in Section~\ref{sect:open_problems}.
Finally, we present an example of use in Section~\ref{sect:example_of_use} and briefly conclude in Section~\ref{sect:conclusion}.

\section{Constraint Solver Approach}
\label{sect:constraint_solver_approach}
There are three constraint-based functional aspects of our approach, all of which we implemented using CHR~\cite{abdennadher.schuetz1998,fruehwirth2009}, based on the CHR-compiler included in SWI-Prolog:

\begin{itemize}
\item path construction,
\item satisfiability checking, and
\item constraint solving.
\end{itemize}

Due to the issue of infeasible paths, at least path construction and satisfiability checking need to be combined in order to facilitate detection of infeasible paths early during path construction.
However, the coupling can be loose in that the path construction phase may be implemented as a separate CHR solver in which the constraints handled by the satisfiability checker are modelled as builtin constraints.

The constraints to be handled represent all relevant expressions and conditions from the underlying language, also covering the different arithmetic types, typically consisting of bounded integer and floating point types.
Information has to be propagated bidirectionally across the borders of these arithmetic theories, for example, when integer values are converted to floating point values or vice versa.

Control-flow is completely represented in the structure of the control-flow graph.
As such, control-flow is current only considered in the path construction module, but not in any of the other constraint solving modules.

The same is true for boolean operations such as conjunction, disjunction and negation.
Most if not all practically relevant programming languages define the evaluation of boolean constructs such that whenever the value of the first operand of a boolean operator completely determines the result, the second operand is not evaluated.
For example, in the expression \verb!A && B!, the second operand \verb!B! is not evaluated if the first, \verb!A! evaluates to \verb!false!, as in this case the result of the whole expression is already determined to be \verb!false!.

In code generation, this is achieved by so-called short-circuit code, which is also applied in our test-data generator.

Because interaction between arithmetic domains may be necessary in presence of type conversions, all arithmetic constraints are handled by a combined constraint handler.

Currently, floating-point constraints are handled by interval filtering~\cite{botella.gotlieb.ea2006,michel2002}.

\subsection{Path Construction}
Our approach to path construction has been previously described in detail~\cite{gerlich2010} and shall be only briefly revisited here.

The path construction approach is centered around a constraint of the form \verb!path(In,A,B,Out)!, meaning: There is a path from node \verb!A! to \verb!B! transforming memory state \verb!In! to memory state \verb!Out!.
The constraint is limited in that \verb!A! and \verb!B! must be ground.

Construction may proceed in one of several ways:

\begin{itemize}
\item forward construction: Construct the path edge-by-edge in the normal direction of execution, starting at \verb!A!.
\item backward construction: Construct the path edge-by-edge in reverse direction, starting at \verb!B!.
\item splitting: Given a node \verb!C! that is (topologically) reachable from \verb!A! and from which \verb!B! can be reached, split up the path from \verb!A! to \verb!B! into two paths, one from \verb!A! to \verb!C! and one from \verb!C! to \verb!B!.
\end{itemize}

All of the three strategies can be described in terms of CHR simplification rules.
However, non-determinism is present in that there may be more than one candidate for the successor, predecessor or intermediate node, respectively.

In absence of any further information indicating paths which may be more likely to help detecting coding defects, bias must be avoided.
To achieve this, alternatives are selected randomly, but not necessarily according to a uniform distribution to achieve appropriate distribution, e.g., of loop iteration counts and total length of the constructed path.

Experience so far indicates that backward construction is most effective in terms of expected time to a feasible path.

\subsection{Linear Integer Constraints}
Linear integer arithmetic is expressed in equations and inequations of the form $e = 0$ and $e \geq 0$, where $e$ is a Presburger expression of the form $c_0 + \sum_{i=1}^n c_i v_i$ with integer constants $c_i$ and logical variables $v_i$.

For example, the assignment $a=2b+c$ is represented as $a-2b-c=0$ or $-a+2b+c=0$.
Similarly, $2a+b>3c$ is represented as $-1+2a+b-3c \geq 0$.

Over the reals, constraint systems of this form could be solved by a combination of Gaussian elimination for the equations and Fourier-Motzkin-elimination for the inequations.
However, Gaussian elimination requires the existence of the multiplicative inverse and Fourier-Motzkin-elimination requires the compactness of the underlying number set, both of which are not given for the integers.

We therefore use an algorithm known as the Omega Test~\cite{pugh1992}.
In this algorithm first the equations are simplified using suitable parameter substitutions.
For example, the equation $a-2b=0$ is trivially processed by substituting $a=2b$ in all other constraints.

Equations which cannot be transformed trivially this way are modified by introducing suitable parameter substitions.
For example, the equation $3a-2b=0$ is equivalent to $a-2\alpha=0 \wedge b=3\alpha$, with $\alpha \in \mathbb{Z}$.
This can be further simplified by substiting $a=2\alpha$, leading to the properly parameterised solution $a=2\alpha \wedge b=3\alpha$.

Inequations are simplified using variable elimination very similar to Fourier-Motzkin-elimination.
There, a variable $w$ can be eliminated from a pair of inequations of the form $c_0 + \sum_{i=1} c_i v_i < a w$ and $b w < d_0 + \sum_{j=1} d_j v_j$ -- with $a,b>0$ -- by introducing a new inequation $b c_0 + \sum_{i=1} b c_i v_i < a d_0 + \sum_{j=1} a d_j v_j$.

Processing all such pairings in this way, all occurrences of $w$ can be eliminated, and the new set of inequations is equivalent to the original set in terms of satisfiability.

Repeating this process for all variables except for one, the original problem is reduced to two inequations of the form $x_l < x < x_u$.
Selecting a value for $x$ from this range and substituting it back into all inequations eliminates the inequations over $x$ and leaves two inequations $y_l < y < y_u$ with $y$ being the variable eliminated second-to-last.
A solution for the original set of inequations can be found by repeating this process of selection and substition until all free variables are bound.

Unfortunately, this only works for compact number sets, i.e. sets where for any $\alpha < \gamma$ there is a $\beta$ with $\alpha < \beta < \gamma$.
This is not the case for the integers.

As a consequence, neither the equivalence regarding satisfiability nor the process for labelling applies when using Fourier-Motzkin-elimination in its usual form.
It is therefore possible that the range for the last variable $x$ is found to be non-empty, but none of the values from this range are part of a solution.

The Omega-Test~\cite{pugh1992} therefore uses a modification of Fourier-Motzkin, under-approximating the set of solutions on each elimination step.
Now any solution of the new set of inequations can be extended to a solution of the original set of inequations.

However, satisfiability of the original set of inequations does not generally imply satisfiability of the inequations after elimination:
As the approximation step may exclude some solutions, these have to be considered separately and in addition to the usual elimination approach.

Depending on the value of the coefficients $a$ and $b$, this set of solutions removed by approximation may grow arbitrarily large.
However, by carefully selecting the order in which the variables are eliminated, the number of solutions to be considered in addition can be reduced significantly.

The algorithm processing equality constraints can be implemented as an online-solver, processing each constraint when it is added to the goal store.
This is of advantage for path construction, where each new step in the program introduces new constraints and requires a new satisfiability check.

Because the quality of elimination results for inequations depends on the order in which variables are eliminated, this process is at best difficult to implement for online processing.
New constraints -- both equations and inequations -- may impact the order in which variables need to be eliminated to reduce the impact of the approximation.
Ad-hoc reordering may thus be necessary.

\subsection{Non-Linear Integer Constraints}
Non-linear integer constraints are handled by a combination of interval filtering and dynamic linear relaxations~\cite{denmat.gotlieb.ea2007}.

Linear relaxations are approximations of non-linear constraints.
For example, the intervals $x \in \left[x_l;x_u\right]$ and $y \in \left[y_l;y_u\right]$ imply that the inequation $\left(x-x_l\right)\left(y-y_l\right) \geq 0$ is fulfilled.
Expanding the expression on the left hand side and using the relationship $z=xy$, this can be transformed to $z-x_l y-x y_l+x_l y_l \geq 0$, which is a linear inequation in $x$, $y$ and $z$.

In a similar manner, the constraints $- z + x y_l + x_u y - x_u y_l \geq 0$, $- z + x y_u + x_l y - x_l y_u \geq 0$ and $z - x y_u - x_u y + x_u y_u \geq 0$ can be derived.

These constraints overapproximate the set of solutions for $x \in \left[x_l;x_u\right] \wedge y \in \left[y_l;y_u\right] \wedge z=xy$, but they contain more information about the relationship between $x$, $y$ and $z$ than the simple interval constraint $z \in \left[z_l;z_u\right]$, where $z_l$ and $z_u$ are derived from $x_l$, $x_u$, $y_l$ and $y_u$.

Dynamic linear relaxations are updated whenever the source information -- in this case the intervals of $x$ and $y$ -- change.

\section{Application of CHR}
\label{sect:application_of_chr}
As has been discussed in previous sections, the constraint solver required for a constraint-based test data generator has to handle a variety of constraint types from different constraint theories normally considered in isolation.
Also, different handling strategies are required at different stages of the test data generation process:
During path construction, the focus is on satisfiability checking.
As soon as a path which is expected to be feasible with sufficient probability is identified, the focus shifts to selection of an actual solution.

Due to the loose coupling possible by way of a rule-based specification concept, CHR allows almost straight-forward integration of the different constraint handlers and of different approaches for solving constraints from the same theory.

One example is the integration of dedicated solution strategies for linear integer constraints besides the more general domain filtering approach for non-linear integer constraints.

This advantage, however, can only be realised for mostly local propagation and simplification strategies, such as the approach to solving linear equations over the integers or domain filtering.

Other strategies such as the elimination procedure of the Omega-Test for linear integer inequations are highly sequential in nature.
Naturally, for implementation of these strategies, imperative languages -- which CHR is not -- are more suited.

\section{Open Problems}
\label{sect:open_problems}
So far the approach has not yet been used for test data generation on industrial-grade software due to several open issues, mainly lack of scalability of the constraint solvers and missing support for arithmetic constraints over floating point numbers.

Industrial software may be large and contain many interactions between different functions, leading to a large number of constraints being generated for a single test-data generation run.
The constraints may be highly coupled, because many of them refer to a small set of variables, namely those variables representing the input values to the function.
Therefore, the issue of scalability is inherent in the problem itself.

Floating point arithmetic plays an increasing role also in embedded software systems in general and in space control software in particular, as more and more embedded hardware platforms have builtin floating point units, thereby gradually replacing the old fixed-point arithmetic implementations.

The theory of constraints over the floats is strictly separate from the theory of constraints over the reals.
This becomes obvious when comparing the magnitude of the underlying sets of numbers:
While the reals are non-countable infinite, the floats are countable and even finite.
This means that a constraint system over the reals may have a solution while the same system is inconsistent when expressed over the floats, and vice versa.

Also, due to the significand-exponent-representation, all operations are non-linear.
In addition, only elementary arithmetic, remainder and square-root have standardised results.
Others, such as the trigonometric functions, are not standardised, the Table Maker's Dilemma being among the reasons~\cite{goldberg1991}.

Some exact domain-filtering approaches to the solution of floating point constraints exist~\cite{botella.gotlieb.ea2006,michel2002}, but their filtering efficiency is insufficient in many situations.

It is quite conceivable that from a theoretical point of view, these problems do not have efficient solutions or any solution at all.
After all, the problem of analytical test-data generation itself cannot be solved in general, as it can be reduced to the halting problem.

However, in practice not the whole set of theoretically applicable operations and their combinations is used.
For example, inequations such as $\sqrt{x^2+y^2}<10^{-7}$ may be expected to occur much more often in a practical context than $x=x+y \wedge y \neq 0$ -- which, due to rounding, has a floating point solution, but requires much more computation time to solve than the latter with current filtering approaches.

As another example, the issue of missing scalability is inherent in the problem, but only when purely focusing on the theoretical description of the problem.
It is likely that in practice the path taken is not influenced by all arithmetic operations performed.
Thus, lazy evaluation -- i.e. introduction of constraints only if they are part of conditions or one or more of their variables are used directly or indirectly in a condition -- may allow for a considerable reduction of complexity in practice, although clearly it would not directly solve the scalability issue once and for all.

Unfortunately, it is quite difficult to define the actual domain of constraint systems to be expected in practice.

\section{Example of Use}
\label{sect:example_of_use}
The approach has not yet been applied for test data generation in practice on industry-grade source-code, mostly due to the open problems stated in Section~\ref{sect:open_problems}.

However, its basic elements were used in the context of a study on the effectiveness of sourcecode-based random test data generation~\cite{gerlich.gerlich.ea2013}.
In the course of this study, random test inputs were injected into functions found in the sourcecode of the control software of an earth observation satellite using the  tool DCRTT (Dynamic C Random Test Tool) developed and maintained in-house at BSSE.
Notably, that control software had previously gone through the rigorous testing and validation stages typical for mission-critical systems, i.e. systems the failure of which could lead to a loss of the satellite of the complete loss of its functionality.

The code was instrumented by the random test tool to monitor for non-specific indications of failures, such as memory access violations, time outs and similar.
Such indications were seen as hints at faults, each of which had to be verified manually to determine whether there was an actual defect or whether the indication was actually a false positive.

In one case, a memory access failure hinted at a critical defect in code related to the installation of in-flight software updates.
A much simplified and anonymised excerpt of the code is given in Listing~\ref{listing:faulty_code}.
The goal of the function is to store data of a given length in a contiguous block at the next free position in the buffer.
If the data block does not fit anymore at the end of the buffer, it shall be stored at the beginning.

\begin{listing}
\begin{verbatim}
#define MAX_BUFFER_SIZE ...
char buffer[MAX_BUFFER_SIZE];
void store_into_buffer(char* data, unsigned int length) {
  const unsigned int last_entry_start = ..., last_entry_length = ...;
  unsigned int next_entry_start = last_entry_start+last_entry_length;
  unsigned int space_available;

  if ((MAX_BUFFER_SIZE - (length-1u)) < next_entry_start)
    next_entry_start = 0;
    
  space_available = (last_entry_start - next_entry_start) % MAX_BUFFER_SIZE;
  if (space_available >= length)
    memcpy(&buffer[next_entry_start],data,length);
  ...
}
\end{verbatim}
\caption{Relevant portions of the faulty function}
\label{listing:faulty_code}
\end{listing}

This code seems short and simple enough to analyse.
However, the defect is non-obvious and the presence of the remainder-operation in combination with a set of choices introduces complexity.
Manual analysis had led to a suspicion for overflowing the buffer in the call to \verb!memcpy!, but was inconclusive both regarding the validity of the suspicion and the possible extent of the overflow.
Due to the complexity, it was not clear whether the results of the analysis were to be trusted.

The matter was settled by providing the path construction part of the test data generator with the goal to reach the call to \verb!memcpy!, adding the constraint \verb!next_entry_start+length>MAX_BUFFER_SIZE!.
Within a second, a descriptive solution was provided, and that solution indicated the potential for a buffer overflow by one byte, namely when \verb!next_entry-start==MAX_BUFFER_SIZE-length+1!, \verb!last_entry_start>0! and \verb!last_entry_start<length-1!.

In this case, there is not sufficient space at the end of the buffer, but the condition \verb!(MAX_BUFFER_SIZE-(length-1u))<next_entry_start! is false.
Therefore the algorithm fails to reset \verb!next_entry_start! to the start of the buffer, and the call to \verb!memcpy! leads to a buffer overflow by exactly one byte.

Further manual analysis led to the conclusion that the one-byte buffer overflow could lead to corruption of volatile data and non-volatile program memory.
Although there was fallback software present for this case -- the so-called \emph{safe-mode software} -- the satellite could have unexpectedly become unresponsive for at least some significant time frame.

We know that in the same project, static verification tools using abstract interpretation have been used to verify the code.
In principle, these tools should have detected the defect by themselves.
It is still not clear whether the tools failed to flag that defect or whether the message got lost in a large number of messages of possible false positives and was therefore not considered.

As a consequence of our report, the defect was fixed.
Also, the instrumentation of the random testing tool was extended to check for such cases, which has led to detection of further defects of the same kind.

\section{Conclusion}
\label{sect:conclusion}
The experience so far has shown that CHR is well-suited for developing complex constraint solvers based on local solution strategies, as detailed in Section~\ref{sect:application_of_chr}.
The correspondence between declarative and operational semantics is helpful in verification of the constraint solver itself.

However, that correspondence often has to be broken in workarounds whenever global strategies need to be implemented, as is the case for the Omega Test.

Further research is required for solving the open problems regarding scalability and handling of floating point constraints.

\section{Acknowledgements}
BSSE is currently performing research on open aspects of industrial-grade CBTDG, supported by a grant by the German federal government under the grant number 50RA1339.

\bibliographystyle{splncs03}
\bibliography{literature}

\end{document}